\documentclass[a4paper,12pt,draft]{article}

\usepackage[cmex10]{amsmath}
\usepackage{latexsym,amssymb}
\usepackage[mathscr]{eucal}
\usepackage{amsxtra,amscd,amsthm,upref,layout,color}
\usepackage{calc}
\usepackage[draft]{graphicx}

\newcommand{\reference}[1]{%
  \par\noindent\parbox{\hsize}{
    #1
  }%
}
\def\href#1{\relax}

\def\br{{\bf r}}\def\bh{{\bf h}} \def\bp{{\bf p}} 
\def\bm{{\bf m}}\def\bk{{{\bf k}}}\def\astar{{{\bf a}^*}}

\def\begeq{\begin{equation}}
\def\endeq{\end{equation}}
\def\begdis{\begin{displaymath}}
\def\enddis{\end{displaymath}}

\def\cB{{\cal B}}\def\cC{{\cal C}}\def\cD{{\cal D}}
 \def\cJ{{\cal J}} \def\cI{{\cal I}} \def\cF{{\cal F}}\def\cH{{\cal H}}
\def\cQ{{\cal Q}}    \def\cN{{\cal N}}
\def\cL{{\cal L}}  \def\cR{{\cal R}}\def\cS{{\cal S}}\def\cV{{\cal V}} 
\def\cZ{{\cal Z}}\def\cZh{{\hat {\cal Z}}}
\def\cQS1{{\cQ_{\cS_1}}}
\def\cHr{{\cH(\cbn)}}

\def\ie{{\it i.e.}}
\def\ieg{{\it e.g.}}

\def\vd#1{{{\vec{\delta}}_{#1}}}  \def\vcR{{\vec{\cR}}}\def\vcP{{\vec{\cal P}}}

  \def\hj{{{{\hat {\jmath}}}}}

\def\cbn{{\bar{\cN}}}

\def\kt#1{{|{#1}\rangle}}\def\rkt#1{{|{#1})}}

\def\bra#1{{\langle {#1}}}\def\rbra#1{{({#1}}}

\def\tom#1#2#3#4{\xi_{#1}^{#3}\eta_{#1 #2}^{#4}}
     
\begin{document}

\title{The Algebraic Approach to the Phase Problem for Neutron Scattering}

\author{
{{A. Cervellino}}${}^a$ {{and S. Ciccariello}}${}^b$\\[6mm]
\begin{minipage}[t]{3mm}
    ${}^a$\ 
\end{minipage}
\begin{minipage}[t]{0.9\textwidth}
    \begin{flushleft}
    \setlength{\baselineskip}{12pt}
{\slshape{\footnotesize{
Consiglio Nazionale delle Ricerche, Istituto di Cristallografia (IC-CNR) 
}}}\\{\slshape{\footnotesize{
Campus Universitario c/o Dip.to Geomineralogico, Via Orabona 4, I-70125 Bari, Italy.
}}}\\{\slshape{\footnotesize{
E-mail {\upshape{\texttt{antonio.cervellino@ic.cnr.it}}}; Phone 
+39~080~5442624;\\ Fax +39~080~5442591
}}}
\end{flushleft}
\end{minipage}\\[10mm]
\begin{minipage}[t]{3mm}
    ${}^b$\ 
\end{minipage}
\begin{minipage}[t]{0.9\textwidth}
    \begin{flushleft}
    \setlength{\baselineskip}{12pt}
{\slshape{\footnotesize{
Universit\`{a} di Padova, Dipartimento di Fisica ``G. Galilei'' \&{} Sez. INFM
}}}\\{\slshape{\footnotesize{
Via Marzolo 8, I-35131 Padova, Italy.
}}}\\{\slshape{\footnotesize{
E-mail {\upshape{\texttt{salvino.ciccariello@pd.infn.it}}}; Phone +39~049~8277173 
;\\ Fax +39~049~8277102
}}}
\end{flushleft}
\end{minipage}
}


\maketitle                        

\begin{abstract}
The algebraic approach to the phase problem for the case of X-ray scattering 
from an ideal crystal is extended to the case of the neutron scattering, 
overcoming the difficulty related to the non-positivity of the scattering 
density. In this way, it is proven that the atomicity is the crucial assumption 
while the positiveness of the scattering density
only affects the method for searching the basic sets of reflections. 
We also report the algebraic expression of the determinants of the 
Karle-Hauptman matrices generated by the basic sets with the most elongated 
shape along one of the reciprocal crystallographic axes.
\vfill
\noindent Preprint catalog: DFPD 03/Th/40
\end{abstract}

\setlength{\baselineskip}{25pt}

\section{Introduction}

The main crystallographic problem, namely: to determine the electron 
density of an ideal crystal with known chemical composition from its 
X-ray diffraction pattern, is brought to its essence when the atoms 
are assumed to be point-like because the positions of the atoms 
present in the unit cell are the only unknown quantities to be 
determined. The unknowns' number being finite, it appears reasonable that 
the knowledge of the peak intensities relevant to a sufficiently large 
portion of the reciprocal lattice is sufficient to determine the atomic 
positions. In fact, Ott (1927) and Avrami (1938) 
first showed that the atomic positions are the roots of a set of polynomial 
equations determined by an appropriate set of reflection intensities. 
This method of inversion of scattering data is known as the \emph{algebraic 
approach} to the phase problem [Buerger (1960), Hauptman (1991)]. Actually, the 
correct formulation of the algebraic approach is slightly more involved 
[Navaza \& Silva (1979), Silva \& Navaza (1981), Rothbauer (1994)] for 
two reasons. Firstly, the unimodular roots of the system of polynomial 
equations, referred to in the following as 
resolvent system, are the positions of the peaks of the infinitely resolved 
Patterson map [Patterson (1939)]. Secondly, for the general case where some of the 
aforesaid peaks have the same projections along one of the three 
crystallographic axes, the resolvent system has to be determined by a 
more involved procedure than Avrami's. These points have been 
fully clarified in two recent papers [Cervellino \& Ciccariello, (1996) and 
(2001)], referred to as I and II in the following. These papers showed the 
existence of many resolvent systems. In particular, the determination of 
resolvent systems is made possible by the positivity of the scattering 
density, ensured by the fact that we are dealing with X-ray scattering. 
Very briefly, according to the basic paper by Goedkoop (1950), the 
positivity condition allows us to associate to each point of the reciprocal 
space lattice $\cZ^3$ a vector of a finite-dimensional Hilbert space $\cHr$. 
The scalar products of these vectors reproduce the intensities of the full 
diffraction pattern. Each resolvent system is determined by a {\it basic set} 
of reflections denoted by $\cB(\cbn)$,  \ie\ by a {\em simply connected} set of 
$\cbn$ reflections such that the associated vectors are linearly independent 
and form a basis of $\cHr$. The coefficients of the polynomial equations 
of the resolvent system require the knowledge of the peak intensities 
relevant to all the reflections obtained as difference of any two 
reflections of $\cB(\cbn)$. Hence, it is important to select $\cB(\cbn)$ in 
such a way that it is centred on ${\bf 0}$ (the origin of reciprocal space) 
and that its points lie as close as possible to the origin. Only when 
the limiting sphere is large enough to contain one of these sets, 
a resolvent system is known and, after solving it, the atomic 
positions can be determined. The procedure to be followed in order 
to select a basic minimal set of reflections was reported in ref. II, where it 
was also shown how to convert each resolvent system of polynomial 
equations in three variables into a resolvent system of polynomial 
equations in a single variable.

The algebraic approach has been successfully applied to solve the 
structure of some real crystals [Fischer \& Pilz (1997) and Pilz 
\& Fischer (2000)] and it can be implemented to account for experimental 
errors on reflex intensities (Cervellino \& Ciccariello, 1999). As a 
matter of fact, its practical usefulness is severely limited by the fact 
that the degree of the polynomial equations sharply increases with the 
number of the atoms present in the unit cell (Hauptman, 1991). 
On a theoretical ground, the approach looks however quite interesting for its 
rigorous conclusions and its far reaching implications since the algebraic approach 
is intimately related to other classical issues of mathematical-physics 
(see the introductory section of  II).

The aim of this paper is to report on the extension of the algebraic 
approach to the case of neutron scattering. As already mentioned, the 
presently known formulations exploit the 
positiveness of the scattering density, 
a condition generally not fulfilled in the case of neutron scattering due 
to the fact that some atomic species have negative scattering lengths. 
On this basis, one rightly wonders whether the mentioned results - in 
particular the property that the full diffraction pattern can be 
reconstructed from the knowledge of the intensities relevant 
to a finite set of reflections, \ie\ the "difference" set generated 
by a basic set of reflections - do apply to neutron scattering or not.  
We shall show that the answer to this question is affirmative\footnote
{To the authors' knowledge, Navaza \& Navaza (1992) already  gave a positive 
answer to this question. However, these authors explicitly recognized to see no rigorous 
way for demonstrating the reconstruction procedure in three dimensions, that is   
the most interesting point. This difficulty is related to that of singling out a 
basic set of reflections from the set of the observed ones. In this paper we 
overcome this difficulty by showing how the isolation of a basic set of 
reflections can be carried out also in presence of a non-positive 
scattering density. By so doing, we generalize the results obtained in 
papers I and II and based on the positeveness of the scattering density.  
It should also be remarked that the aforesaid proof of the reconstruction property 
requires no probabilistic assumption. Therefore, our conclusions are more general 
than those obatined by Hauptman (1976) with the probabilistic approach that, for 
practical appplications, is by far the most useful one (see, \ieg, Hauptman and Langs (2003)).}. 
In order to prove this 
statement, it is necessary to relax the positiveness condition. Hence, 
the plan of the paper is as follows. In \S 2 we report the basic equations 
of the algebraic approach and the finite vectorial space $\cHr$ will be introduced 
on the basis of simple quantum mechanical notions. Based on the results proven 
in Appendices A and B, in \S 3 we generalize the algebraic approach to neutron scattering 
in the case of two-dimensional crystals and in \S 4 we sketch the generalization 
to the three-dimensional case and report our conclusions. 
Appendix A illustrates a new procedure, not requiring the positivity assumption,  
for singling out a {\em principal} basic set of reflections, where 
{\em principal} means that the basic set has the most elongated shape along one 
axis of reciprocal space. Appendix B deals with the derivation of the algebraic 
expression of the determinant of the Karle-Hauptman matrix associated to a 
principal basic set of reflection. 

\section {Basic results of the algebraic approach}

The formulation of the algebraic approach, reported in I and II, assumed 
positivity. We will now retrace our steps through the theory in order to 
make the necessary changes to allow for non-positive scattering densities, 
as it happens with neutrons. We continue to assume that the unit cell 
contains $N$ point-like atoms. Its scattering density 
has the following expression 
\begin{equation}
\label{1.1}
\rho_{cell}(\br)=\sum_{j=1}^N {\cZh}_j \delta(\br -\br_j)
\end{equation}
where $\delta(\cdot)$ is the three-dimensional (3D) Dirac function, $\br_j$ the 
position of the $j$th atom and $\cZh_j$ the atomic number or the scattering-length 
of the $j$th atom, depending on whether one considers X-ray or neutron scattering. 
The two cases differ because the $\cZh_j$'s are {\it positive} integers in the case 
of X-rays and only real numbers in the case of neutrons\footnote{It is understood 
that absorption and other experimental effects are either absent or corrected for.}.  Thus, the 
positiveness of the scattering density is generally not ensured in the neutron 
case. But $I_{obs,\bh}$ - the intensity observed at reflection $\bh$ - is in both cases 
the square modulus of the Fourier transform of (\ref{1.1}), \ie
\begin{equation}\label{1.2}
I_{obs,\bh}=\biggl |\sum_{j=1}^N {\cZh}_j e^{i2\pi \bh\cdot\br_j}\biggr |^2 
=\sum_{j=1}^N {\cZh}_j^2 + \sum_{1\le j\ne k\le N}{\cZh}_j\cZh_j e^{i2\pi \bh\cdot(\br_j-\br_k)}.
\end{equation}
Each vector $(\br_j-\br_k)$ can be brought within the unit cell by adding to it a 
vector ${\bm}_{j,k}$ with components equal to 0 or -1, so as to write
\begin{equation}\label{1.3}
\br_j-\br_k + \bm_{j,k} ={\vec \delta}.
\end{equation}
As $(j,k)$ runs over its $N(N-1)$ values, we label the different ${\vec \delta}$'s, defined 
by (\ref{1.3}), by $\hj$ and we  denote by $\cbn'$ the number of the different $\vd{\hj}$'s. 
Moreover, we denote by $\cL_{\hj}$ the set of pairs $(j,k)$ such that $(\br_j-\br_k)$ 
defines the same $\vd{\hj}$ after applying (\ref{1.3}). Then, the second sum on the 
right hand side (rhs) of (\ref{1.2}) becomes 
\begin{equation}
\sum_{\hj=1}^{\cbn'}e^{i2\pi \bh\cdot\vd{\hj}}\sum_{(j,k)\in \cL_{\hj}}\cZh_j\cZh_k
\end{equation}
After setting 
\begin{equation}\label{1.4a}
\nu_{\hj}\equiv \sum_{(j,k)\in \cL_{\hj}}\cZh_j\cZh_k,
\end{equation}
and 
\begin{equation}\label{1.4b}
I_{\bh}\equiv I_{obs,\bh} - \sum_{j=1}^N {\cZh}_j^2,
\end{equation}
Eq.(\ref{1.2}) reads
\begin{equation}\label{1.5}
I_{\bh} = \sum_{\hj=1}^{\cbn}\nu_{\hj}e^{i2\pi \bh\cdot\vd{\hj}},
\end{equation}
where $\cbn$ is the number of the $\nu_{\hj}$'s different from zero. 
[In the case of neutron scattering, $\cbn$ can be smaller than $\cbn'$ because 
the negativeness of some $\cZh_j$'s can make some $\nu_{\hj}$'s equal to zero.]
Eq.(\ref{1.5}) shows that the $I_{\bh}$'s, the "subtracted" peak intensities defined 
by Eq. (\ref{1.4b}), are the Fourier transforms of the scattering density 
relevant to the Patterson map 
\begin{equation}\label{1.6}
\rho_{_{Pat}}(\br)=\sum_{\hj=1}^{\cbn}\nu_{\hj}\delta(\br-\vd{\hj}). 
\end{equation}
This consists of $\cbn$ scattering centres located at ${\vec\delta}_1,\dots,
{\vec\delta}_{\cbn}$ with weights or "charges" equal to $\nu_1,\dots,\nu_{\cbn}$, and 
the positiveness of the weights is ensured only in the case of X-ray scattering. Moreover, 
Eqs.~(\ref{1.4b}), (\ref{1.5}) and (\ref{1.6}) make it evident that the knowledge of all 
the observed intensities $I_{obs,\bh}$ only determines quantities $\vd{\hj}$'s and 
$\nu_{\hj}$'s, \ie\ the scattering density of the Patterson map. Assuming the latter 
quantities known, Eqs~(\ref{1.3}) and (\ref{1.4a}) can be inverted to determine the 
atoms' positions $\br_1,\ldots,\br_N$ by the procedure reported in \S 3.2 of I. 
This deconvolution of the Patterson map involves a finite number of operations. In this 
way, all the atomic configurations that reproduce the observed diffraction pattern are 
determined. Hence, the difficult problem to be solved is to find out the set of 
Eq.s~(\ref{1.5}) that uniquely determine $\cbn$, $\vd{\hj}$ and $\nu_{\hj}$ for $\hj=1,
\dots,\cbn$. The solution of this problem requires, firstly, the choice of an appropriate 
set of $\bh$ values that determine the equations to be solved and, secondly, a procedure 
able to solve the resulting set of non-linear equations. 

For X-ray scattering, the solution of the first problem is achieved by introducing the 
Goedkoop (1950) lattice of vectors, which is a subset of a finite-dimensional Hilbert 
space. Unfortunately, this step requires that all the $\nu_{\hj}$'s are positive and, 
therefore, it cannot be extended to the case of neutrons. However, by using some notions 
of elementary Quantum Mechanics, we show now that in both cases it is possible to introduce 
a finite-dimensional Hilbert space and, within the latter, a lattice of vectors in such 
a way that the scattering density (\ref{1.6}) and the "subtracted" intensities 
(\ref{1.5}) are two different representations of a single hermitian operator.

To this aim we recall that the position and momentum operator, respectively denoted by 
$\vcR$ and $\vcP$, have eigenvectors $\kt{\br}$ and $\kt{\bp}$ whose eigenvalues $\br$ and 
$\bp$ span the full 3D space $R^3$. Consider now the eigenvalues ${\bp}$ equal 
to $-2\pi{\bh}$, $\bh$ being a triple of integers, and put $\rkt{\bh}\equiv\kt{-2\pi{\bh}}$. 
As $\bh$ ranges over the 3D lattice $\cZ^3$, the set of $\rkt{\bh}$'s defines a lattice 
of vectors lying within the infinite-dimensional Hilbert space $\cH$ spanned by the 
eigenvectors $\kt{\bp}$ or $\kt{\br}$.  Introduce now the linear operator
\begin{equation}\label{1.7}
{\cQ}\equiv \sum_{\hj=1}^{\cbn}\kt{\vd{\hj}} \nu_{\hj}\bra{\vd{\hj}}|,
\end{equation}
where $\kt{\vd{\hj}}$ is the eigenvetor of $\vcR$ with eigenvalue $\vd{\hj}$ equal to the 
position vector of the $\hj$th scattering centre. Due to the property 
$\bra{\br}\kt{\br'}=\delta(\br-\br')$, the matrix elements of $\cQ$ with respect to the 
eigenvectors of $\vcR$ are
\begin{equation}\label{1.8a}
\bra{\br}|\cQ\kt{\br'}=\delta(\br-\br')\sum_{\hj=1}^{\cbn}\nu_{\hj}\delta(\br-\vd{\hj})
\end{equation}
At the same time, the matrix elements of $\cQ$ with respect to the lattice vectors 
$\rkt{\bh}$ are 
\begin{equation}\label{1.8b}
\rbra{\bh}|\cQ\rkt{\bh'}=(2\pi)^{-3}\sum_{\hj=1}^{\cbn}\nu_{\hj}e^{i2\pi\vd{\hj}\cdot(\bh-\bh')},
\end{equation}
where we used the property that $\bra{\bp}\kt{\br}=e^{i{\bp}\cdot\br}/(2\pi)^{3/2}$ and 
units such that $\hbar=1$ (Messiah, 1959). Comparison of (\ref{1.8a}) with (\ref{1.6}) 
shows that the scattering density (\ref{1.6}) coincides with the diagonal matrix elements 
of $\cQ$ (leaving aside the divergent factor related to the value $\delta({\bf 0})$ of 
the 1st Dirac function). On the other hand, the comparison of (\ref{1.8b}) with (\ref{1.5}) 
shows that all the subtracted intensities (\ref{1.5}) are $(2\pi)^3$ times the matrix 
elements of $\cQ$ with respect to the lattice vectors $\rkt{\bh}$. Moreover, Eq.(\ref{1.7}) 
shows that the "charge density" operator $\cQ$ is determined only by the $\cbn$ eigenvectors 
$\kt{\vd{1}},\ldots,\kt{\vd{\cbn}}$ of $\vcR$ with eigenvalues equal to the positions of 
the $\cbn$ scattering centres, and by the $\cbn$ real numbers $\nu_1,\ldots,\nu_{\cbn}$ 
equal to the weights of the scattering centres. Hence, we can restrict ourselves to the 
finite-dimensional Hilbert space $\cHr$ spanned by the vectors $\kt{\vd{1}},\ldots,
\kt{\vd{\cbn}}$ and defined as 
\begdis
\cHr\equiv \Bigl\{ \kt{v}=\sum_{\hj=1}^{\cbn} \alpha_{\hj}\kt{\vd{\hj}}\Bigl |\ \ \alpha_{1},
\ldots,\alpha_{\cbn}\in C\Bigr \}.
\enddis  
Vectors $\kt{\vd{1}},\ldots,\kt{\vd{\cbn}}$ obey the orthonormality condition\footnote{By 
so doing,the previous normalization 
$\bra{\vd{\hj'}}\kt{\vd{\hj}}=\delta(\vd{\hj'}-\vd{\hj})$ 
has been scaled to  $\bra{\vd{\hj'}}\kt{\vd{\hj}}=\delta_{\hj',\hj}$.}
\begin{equation}\label{1.9a}
\bra{\vd{\hj'}}\kt{\vd{\hj}}=\delta_{\hj',\hj},\quad  \hj,\hj'=1,\dots,\cbn,
\end{equation}
$\delta_{\hj',\hj}$ being the Kronecker symbol, as well as the completeness relation 
\begin{equation}\label{1.9b}
\sum_{\hj=1}^{\cbn}\kt{\vd{\hj}}\bra{\vd{\hj}}| = 1. 
\end{equation}
In order to preserve the validity of (\ref{1.8b}), we still need to assume that $\cHr$ contains 
a lattice of vectors $\kt{\bh}$ [not to be confused with $\rkt{\bh}$ or with the eigenvectors 
of $\vcP$, see the following Eq.~(\ref{1.10d})]. To his aim, it is sufficient to put 
\begin{equation}\label{1.10b}
\kt{\bh}\equiv 
\sum_{\hj=1}^{\cbn}e^{-i2\pi\bh\cdot\vd{\hj}}\kt{\vd{\hj}},\quad\forall \bh\in\cZ^3.
\end{equation}
After taking the scalar product with $\bra{\vd{\hj'}}|$ one gets 
\begin{equation}\label{1.10a}
\bra{\vd{\hj'}}\kt{\bh}=e^{-i2\pi\bh\cdot\vd{\hj'}},\quad \forall \bh\in\cZ^3,\ {\hj}'
={1,2,\ldots,\cbn}.
\end{equation}
>From the above two relations it follows that vectors $\kt{\bh}$ are no longer orthogonal 
since from (\ref{1.10b}) and (\ref{1.9a}) one gets 
\begin{equation}\label{1.10c}
\bra{\bh}\kt{\bh'}=\sum_{\hj=1}^{\cbn}e^{i2\pi\vd{\hj}\cdot(\bh-\bh')}=\bra{\bh+\bm}
\kt{\bh'+\bm},\quad{\forall}\ \bh,\bh',\bm\in\cZ^3,
\end{equation}
with $\bra{\bh}\kt{\bh}=\cbn$. This property is not surprising if one observes that 
$\kt{\bh}$ and $\rkt{\bh}$ are related as follows
\begin{equation}\label{1.10d}
\kt{\bh}=(2\pi)^{3/2}\sum_{\hj=1}^{\cbn}\kt{\vd{\hj}}\bra{\vd{\hj}}\rkt{\bh}
\end{equation}
so that $\kt{\bh}$ is the projection of $\rkt{\bh}(\in \cH)$ into $\cHr$ and, therefore, 
it is no longer an eigenvector of $\vcP$. Now it is important to note that 
\begin{equation}\label{1.11}
\bra{\bh'}|\cQ\kt{\bh}=\sum_{\hj=1}^{\cbn}\nu_{\hj}e^{i2\pi\vd{\hj}\cdot(\bh'-\bh)}=I_{\bh'-\bh}.
\end{equation}
Thus, on the one hand, all the matrix elements of $\cQ$ with respect to the lattice of 
vectors $\kt{\bh}$ reproduce the full diffraction pattern. On the other hand, the 
diagonal matrix elements of $\cQ$ with respect to the basis vectors $\kt{\vd{\hj}}$ 
are the weights of the scattering density (\ref{1.6}). In this way, it has been 
shown that: i) both for X-ray and for neutron scattering it can be introduced 
a finite-dimensional Hilbert space $\cHr$ spanned by the $\cbn$ eigenvectors of 
$\vcR$ associated to the position vectors of the $\cbn$ scattering centres, ii) 
within $\cHr$ it exists a lattice of vectors $\cZ^3_v\equiv\ \{\kt{\bh}\,|\ 
\bh\in \cZ^3\}$ with $\kt{\bh}$ defined by Eq.~(\ref{1.10b}), iii) it exists 
a hermitian linear operator $\cQ$ whose matrix elements with respect to the 
basis vectors $\kt{\vd{\hj}}$ and to the vectors of the vectorial lattice $\cZ^3_v$ 
yield all the weights of the scattering density and all the subtracted intensities 
$I_{\bh}$, respectively. 

We are now left with the problem of determining $\cbn$, the $\kt{\vd{\hj}}$'s and 
the $\nu_{\hj}$'s knowning an appropriate number of $I_{\bh}$ values. Before 
tackling with this problem in the following sections, we report some interesting 
consequences of the aforesaid vectorial structure underlying the phase problem 
for an ideal crystal. The first, related to Eq.~(\ref{1.10c}), shows that 
$\bra{\bh'}\kt{\bh}=\bra{\bh'+\bm}\kt{\bh+\bm},\ \forall \bh,{\bh'},\bm\in \cZ^3$. 
Thus, the scalar product of any two vectors of $\cZ^3_v$ does not change if 
the associated lattice points are translated by an arbitray vector $\bm\in \cZ^3$. 
The same property applies to $\bra{\bh'}|\cQ\kt{\bh}$ because by Eq.~(\ref{1.11}) 
one gets 
\begin{equation}\label{1.12a}
I_{\bh'-\bh}=\bra{\bh'}|\cQ\kt{\bh}=
\bra{\bh'+\bm}|\cQ\kt{\bh+\bm}=\bra{\bh'-\bh}|\cQ\kt{{\bf 0}},\  \forall\ \bh, 
\bh', \bm\in \cZ^3.
\end{equation}
Second, $\cZ^3_v$ cannot contain more than $\cbn$ linearly independent vectors 
because it is a subset of $\cHr$. In Appendix A it will be shown that $\cZ^3_v$ 
exactly contains $\cbn$ linearly independent vectors. Thus, if we denote one set 
of these vectors by $\kt{\bk_1},\kt{\bk_2},\dots,\kt{\bk_{\cbn}}$, we can write
\begin{equation}\label{1.12}
\kt{\bh}=\sum_{j=1}^{\cbn}A_{\bh,j}\kt{\bk_j},\quad \forall \bh\in \cZ^3.
\end{equation}
Taking the scalar product of the adjoint of this equation with vector $\kt{\bf 0}$ 
and using Eq.~(\ref{1.12a})  one obtains 
\begin{equation}\label{1.13a}
I_{\bh}=\sum_{j=1}^{\cbn}{\bar A}_{\bh,j}\bra{\bk_j}|\cQ\kt{{\bf 0}}=
\sum_{j=1}^{\cbn}{\bar A}_{\bh,j}I_{\bk_j}\quad\forall \bh\in \cZ^3,
\end{equation}
where the overbar denotes the complex conjugate. Eq.~(\ref{1.13a}) shows that any 
subtracted intensity $I_{\bh}$ is a linear combination of the $\cbn$ intensities 
$I_{\bk_1},\ldots,I_{\bk_{\cbn}}$. The matrix elemets $A_{\bh,j}$ obey to a set of 
relations. The first is obtained by taking the scalar product of (\ref{1.12}) with 
$\bra{\vd{\hj}}|$ and reads 
\begin{equation}\label{1.14a}
e^{-i2\pi\bh\cdot\vd{\hj}}=\sum_{\ell=1}^{\cbn}A_{\bh,\ell}e^{-i2\pi\bk_{\ell}\cdot
\vd{\hj}},\quad \forall\bh\in\cZ^3,\  \hj={1,2,\cdots,\cbn}.
\end{equation}
After introducing an $\cbn\times\cbn$ matrix $(\cV)$ with $\cV_{\hj,\ell}\equiv 
e^{-i2\pi\bk_{\ell}\cdot\vd{\hj}}$, the previous equations becomes 
\begin{equation}\label{1.14b} 
\sum_{\ell=1}^{\cbn}\cV_{\hj,\ell}A_{\bh,\ell}=e^{-i2\pi\bh\cdot\vd{\hj}},
\end{equation}
which, as it will be shown later, can formally be solved as 
\begin{equation}\label{1.14c}
A_{\bh,\ell}=\sum_{\hj=1}^{\cbn}(\cV^{-1})_{\ell,\hj}e^{-i2\pi\bh\cdot\vd{\hj}}
\end{equation}
The second is obtained by substituting Eq.~(\ref{1.12}) in Eq.~(\ref{1.12a}), 
using Eq.~(\ref{1.13a}) and the linear independence of $\kt{\bk_1},\ldots,\kt{\bk_{\cbn}}$. 
One finds 
\begin{equation}\label{1.14d}
{\bar A}_{\bh-\bh',l}=\sum_{j=1}^{\cbn} \sum_{j'=1}^{\cbn}{\bar A}_{\bh,j}A_{\bh',j'}
{\bar A}_{\bk_j-\bk_{j'},l}\quad \forall \bh,\bh'\in {\cZ}^3,\ l={1,2,\ldots,\cbn}.
\end{equation}
\bigskip
\section{Neutron scattering}
Given their lengthy nature, we report in Appendix A the proof of the existence of 
the principal basic set of vectors $\kt{\bh}$ along a crystallographic axis, and 
in Appendix B the analytic expression of the determinant of the Karle-Hauptman 
matrix associated to this principal basic set of vectors. It is stressed that 
these results, worked out in the two-dimensional 
case $(D=2)$, apply both for X-ray and for neutron scattering, because we 
never require the posivity in proving them. Moreover, they easily generalize 
to the three-dimensional case.  In this section we show how to single out 
a basic set of reflections in the case of neutron scattering,  
knowing an appropriate number of subtracted intensities 
$I_{\bh}$. [For simplicity we shall still confine ourselves to the case $D=2$.] 
Firstly, we recall that the results, obtained in the previous section as well 
as in Appendices A and B independently of the positeveness, show that:  
i) it is possible to introduce a Hilbert space $\cHr$ having as an orthonormal 
basis the set of the (scaled) eigenvectors $\kt{\vd{\hj}}$ of $\vcR$ with eigenvalues 
equal to the position vectors $\vd{\hj}$ of the $\cbn$ scattering centres of the 
(infinitely resolved) Patterson map determined by the full diffraction pattern, 
ii) within $\cHr$ it is possible to extract a lattice $\cZ^2_v$ of vectors 
$\kt{\bh}$ defined by (\ref{1.10b}), iii) the (subtracted) intensities defined 
by (\ref{1.4b}) are the matrix elements of the charge density operator $\cQ$ 
[defined by (\ref{1.7})] with respect to vectors of $\cZ^2_v$, while the charges 
$\nu_{\hj}$ are the eigenvalues of the eigenvectors $\kt{\vd{\hj}}$ of 
$\cQ$, iv) the knowledge of the eigenvalues $\vd{\hj}$ allows us to determine the subset 
$\cI$ [defined by (\ref{lattb})] of $\cZ^2$  and the translation of $\cI$ by 
$(-1,-1)$ yields the principal basic set $\cB(\astar)$ of $\cHr$ along the 
reciprocal crystallographic direction $\astar$ and, finally, 
v) the algebraic expression of the determinant of the KH matrix associated to 
$\cB(\astar)$ is given by Eq.~(\ref{detkh}). Even though results i)-iv) were 
essentially obtained in papers I and II, it is stressed that they are 
now extended to the case of neutrons. Moreover, the above presentation makes the 
introduction of $\cHr$ clearer and shows that the $\nu_{\hj}$'s and the $I_{\bh}$'s 
are the matrix elements of a single operator $(\cQ)$ with respect to two different 
sets of vectors. As yet, however, the aforesaid generalization is practically useless, 
because we do not know the $\vd{\hj}$'s  and $\cbn$. Since the only known quantities 
are the subtracted intensities $I_{\bh}$, the determination of the $\vd{\hj}$'s, 
$\nu_{\hj}$'s and $\cbn$ must be carried out in terms of the $I_{\bh}$'s. Hence, 
the search of a basic set must be performed in terms of these quantities. 
In papers I and II, we reported the procedures for carrying through such a search. 
Unfortunately, they only apply to the case of X-ray scattering because they exploit 
the positiveness of the charge density operator $\cQ$, a condition fulfilled only in 
the case of X-rays. Very briefly, as shown in I and II, the simplest 
search of a basic set proceeds as follows. One starts from the set of reflections 
$\cB_2=\{(0,0), (1,0)\}$ and one evaluates the determinant of the associated KH matrix 
$\bigl(\cD[\cB_2]\bigr)$. If $\det\Bigl(\cD[\cB_2]\Bigr)\ne 0$, one "enlarges" $\cB_2$ 
by "adding" to it the next reflection (2,0) so as to have the new set $\cB_3=\{(0,0), 
(1,0),(2,0)\}$. Then one evaluates $\det\Bigl(\cD[\cB_3]\Bigr)$ and if this value is 
different from zero one adds the next reflection (3,0) to $\cB_3$ so to have the enlarged 
set $\cB_4$. As far as the vectors associated to the reflections of the considered set are 
linearly independent, the positiveness of $\cQ$ ensures that the corresponding KH matrix 
has a strictly positive determinant. On the contrary, if the vectors are linearly 
dependent the determinant is equal to zero. This property is easily shown 
as follows. Assume that $\cB_m=\{\bh_1, \bh_2,\ldots,\bh_m\}$ and consider the 
$m\times m$ matrix $(\cQ_m)$ having its $(r,s)$ element given by 
\begin{equation}\label{5.1}
\bra{\bh_r}|\cQ\kt{\bh_s}=I_{\bh_r-\bh_s}=\sum_{\hj=1}^{\cbn}\bra{\bh_r}\kt{\vd{\hj}}
\nu_{\hj}\bra{\vd{\hj}}\kt{\bh_s}=\sum_{\hj=1}^{\cbn}{\overline \cV_{\hj,r}}\nu_{\hj}
\cV_{\hj,s},\quad{r,s=1,\dots,m}
\end{equation}
where Eq.~(\ref{1.10a}) and the definition of $\cV_{\hj,l}$ reported below 
Eq.~(\ref{1.14a}) have been used. Eqs.(\ref{5.1}) can be written in matricial form as 
\begin{equation}\label{5.2}
\Bigl(\cQ_m\Bigr) = \Bigl(\cV^{\dag}\Bigr)\cdot \Bigl(v\Bigr) \cdot\Bigl(\cV\Bigr)
\end{equation}
where $\Bigl(\cV\Bigr)$ is now an $\cbn\times m$ matrix (with $m<\cbn$), 
$\Bigl(\cV^{\dag}\Bigr)$ its hermitan conjugate and $(v)$ an 
$\cbn\times \cbn$ diagonal matrix with its $(\hj,\hj')$ element equal to $\nu_{\hj}
\delta_{\hj,\hj'}$. The determinant of matrix $(\cQ_m)$,  
evaluated by Bezout's theorem (Gantmacher, 1966), yields 
\begin{equation}\label{5.3}
\det(\cQ_m)=\sum_{1\le {\hj}_1<{\hj}_2<\dots<{\hj}_m\le \cbn}\bigl
[\nu_{{{\hj}_1}}\dots\nu_{{{\hj}_m}}\bigr ]\bigl | 
\det(\cV_{{\hj}_1,\dots,{\hj}_m})\bigr |^2,
\end{equation} 
where $(\cV_{{\hj}_1,\dots,{\hj}_m})$ denotes the $m\times m$ minor formed with the 
$\hj_1$th,...,$\hj_m$th row of $(\cV)$. In the case of X-rays the non-negativeness of 
the addends present in the above sum implies that $\det(\cQ_m)\ne 0$ unless all the quantities 
$\det(\cV_{{\hj}_1,\dots,{\hj}_m})$ are equal to zero for $1\le {\hj}_1<{\hj}_2
<\dots<{\hj}_m\le \cbn$.  The latter conditions are verified if and only if the 
considered vectors $\kt{\bh_1},\ldots,\kt{\bh_m}$ are linearly dependent. In fact, 
when this condition is fulfilled, the generic vector
\begdis
\kt{a}=\sum_{r=1}^m\alpha_{r}\kt{\bh_r},\quad \alpha_1,\ldots,\alpha_m\in {\rm C}
\enddis
can be equal to zero with some $\alpha_j$'s different from zero. 
Then, the scalar products with $\bra{\vd{\hj}}|$ yields 
\begdis
\bra{\vd{\hj}}\kt{a}=\sum_{r=1}^m\alpha_{r}\bra{\vd{\hj}}\kt{\bh_r}=\sum_{r=1}^m
\alpha_{r}\cV_{\hj,r}=0, \quad \hj=1,\ldots,\cbn.
\enddis
This being a set of $\cbn$ linear homogeneous equations in the unknown $\alpha_{1},
\ldots,\alpha_{m}$, the Rouch\`e-Capelli theorem ensures that a non-trivial solution 
exists if and only if the rank of the $\cbn\times m$ matrix $(\cV)$ is smaller than 
$m$, \ie\ if the determinants of all the $m\times m$ matrices contained in $(\cV)$ are equal 
to zero. In this case, one finds that $\det(\cQ_m)=0$ and the vectors of the 
considered set $\cB_m$ are linearly dependent. On the contrary, if one of the 
$m\times m$ matrix contained in $(\cV)$ is non singular, all the $\alpha_m$'s are 
equal to zero, the vectors $\kt{\bh_1},\ldots,\kt{\bh_m}$ are linearly independent 
and $\det(\cQ_m)\ne 0$. Coming back to the search procedure of a basic set, the 
aforesaid property makes it clear that the enlargement procedure comes to a halt 
when $\det(\cD[\cB_{\bar m}])=0$, \ie\ when one finds a "KH zero". This condition 
must certainly occurr because the number of the linearly independent vectors cannot 
exceed $\cbn$. Hence, according to the analysis reported in Appendix A, we find that 
${\bar m}=\mu_1=M$ and the vectors associated to the set $\cB_{\mu_1}=\{(0,0),\ldots,(M-1,0)\}$ 
are linearly independent. Then, we enlarge the previous set by adding to it, step by 
step, the reflections lying on the next upper row starting with $(0,1)$. 
The next KH zero is found when we "add" the reflection $(\mu_2,1)$. We move now 
to the next upper row and we start by adding the reflection (0,2) to the set 
$\cB_{\mu_1,\mu_2}=\{(0,0),\ldots,(\mu_1-1,0),(0,1),\ldots,(\mu_2-1,1)\}$. The search 
of the basis set is accomplished when we arrive at the point where the inclusion of 
the reflection $(0,m_1)$ leads to a KH zero. By construction, the resulting basis set 
is simply connected. 

In the case of neutron scattering, this analysis is no longer possible because the 
finding of a KH zero (\ie\ $\det(\cQ_m)=0$), during the enlargement procedure, does 
not allow us to infer that the associated vectors are linearly dependent. This 
appears evident from Eq.~(\ref{5.3}): here, each factor related to the charge product 
is not ensured to be positive so that the condition $\det(\cQ_m)=0$ does not imply 
that all the quantities $\bigl | \det(\cV_{{\hj}_1,\dots,{\hj}_m})\bigr |^2$ are equal 
to zero. Thus, in order to extend the search procedure of a basic set to the case of 
neutron scattering, we must introduce a positive definite operator whose matrix 
elements with respect to the vectorial lattice $\cZ^2_v$ are known in terms of the 
observed scaled intensities $I_{\bh}$. To this aim, denote by $\cS_{obs}$ the set of 
the observed reflections and denote by $\cS_1$ the largest subset of $\cS_{obs}$ such 
that for any two reflections $\bh_r$ and $\bh_{r'}$ in $\cS_1$ it results $(\bh_r -
\bh_{r'})\in \cS_{obs}$. We denote by $\cbn_1$ the number of reflections contained 
in $\cS_1$ and we assume first that $\cS_1$ is large enough to contain at least 
one basic set so that $\cbn_1>\cbn$. Consider now the linear operator\footnote{The 
introduction of this quantity is suggested by the procedure followed by Silva \& 
Navaza (1981) and Navaza \& Navaza (1992).}
\begin{equation}\label{5.4}
\cQS1\equiv \cQ \sum_{r=1}^{\cbn_1}\kt{\bh_r}\bra{\bh_r}|\cQ. 
\end{equation}
This operator is hermitian and positive definite. The first property is evident. To 
show the second, we consider the expectation value of $\cQS1$ with respect to an 
arbitrary vector $\kt{a}\in \cHr$. One finds that 
\begdis
\bra{a}|\cQS1\kt{a}=\sum_{r-1}^{\cbn_1}\bigl |\bra{a}|\cQ\kt{\bh_r}\bigl|^2
\enddis
This expectation value can be equal to zero either if $\cQ\kt{a}$ is perpendicular to 
all the $\kt{\bh_r}$'s for $r=1,\ldots,\cbn_1$ or if $\cQ\kt{a}=0$. The first condition 
is impossible unless $\cQ\kt{a}=0$ because $\cS_1$ is assumed to contain a basic set of 
vectors. We are left with the condition $\cQ\kt{a}=0$. This implies that $\kt{a}$ 
is eigenvector of $\cQ$ with eigenvalue 0. But this condition is impossible 
because the eigenvalues of $\cQ$ are all different from zero as it appears evident from 
Eq.~(\ref{1.7}). Hence, $\bra{a}|\cQS1\kt{a}>0$ \ $\forall\kt{a}\ne 0$ and the positivity 
of $\cQS1$ is proven. The matrix elements of $\cQS1$ with respect to the vectors of the 
lattice $\cZ^2_v$ are 
\begdis
\bra{\bh}|\cQS1\kt{\bk}=\sum_{r=1}^{\cbn_1} I_{\bh-\bh_r}I_{\bh_r-\bk}. 
\enddis
If $\bh,\bk\in \cS_1$, the matrix elements of $\cQS1$ are fully known and will be denoted 
as 
\begin{equation}\label{5.5}
\cJ_{\bh_l,\bh_m}\equiv \bra{\bh_l}|\cQS1\kt{\bh_m}=\sum_{r=1}^{\cbn_1} I_{\bh_l-\bh_r}
I_{\bh_r-\bh_m}, \quad l,m=1,\ldots,\cbn_1.
\end{equation}
[It is noted that the $\cJ_{\bh_l,\bh_m}$'s are symmetric since they obey the relation 
$\cJ_{\bh_l,\bh_m}=\cJ_{\bh_m,\bh_l}$ that follows from the Friedel property valid for 
the subtracted intensities, \ie\ $I_{\bh}=I_{-\bh}$.] At this point, the search of a 
basic set becomes possible acting as follows. We start from the set of reflections 
$\cB_2=\{(0,0),(0,1)\}$ and we evaluate the determinant of the matrix whose elements are 
the matrix elements of $\cQS1$ between the vectors associated to $\cB_2$. 
These matrix elements are known owing to (\ref{5.5}). For 
simplicity, this matrix also will be called a Karle-Hauptman matrix, even though its 
matrix elements are the $\cI_{\bh_l,\bh_m}$'s instead of the $I_{\bh_l,\bh_m}$'s. 
If the determinant of this KH matrix is different from zero, we enlarge $\cB_2$ by 
adding to it a further reflection chosen 
either by the procedure described above or by one of the other procedures reported in II. 
Assume that $\cB_m$ is the first set, found during the enlargement procedure, such that 
the determinant of the KH matrix (with elements $\cI_{\bh_l,\bh_m}$) is equal to 
zero. Since $\cQS1$ is a positive definite operator, the vectors $\kt{\bh_1}$,..,$
\kt{\bh_m}$  associated to the reflections of $\cB_m$ are linearly dependent. Moreover,  
the vectors $\kt{\bh}$ considered in this paper do not depend on the charges $\nu_{\hj}$ [see 
Eq.~(\ref{1.10b})]. Actually, as explained in footnote 6, they refer to positive charges. 
Thus, property 1 proven in sect. 4 of II applies. Therefore, in the subsequent enlargement 
procedure, we must discard all the reflections relevant to the quadrant defined in property 1. 
The search of a basic  set will be accomplished once the resulting set cannot further 
be enlarged by the adopted procedure of enlargement. 
If this happens, one concludes that the considered $\cS_1$ set and, consequently, 
the underlying limiting sphere are large enough to contain a basic set. Before 
discussing the consequences of this result, we need to say what happens when $\cS_1$ 
is not large enough to contain a basic set of vectors. In this case, we can always 
denote by $\kt{\bh_1},\ldots,\kt{\bh_M}$  the $M(<\cbn_1)$  linearly independent 
vectors contained in $\cS_1$ that are closer to the origin of reciprocal space. 
The remaining vectors 
$\kt{\bh_{M+1}},\ldots,\kt{\bh_{\cbn_1}}$ of $\cS_1$ take the form 
\begdis
\kt{\bh_r}=\sum_{r=1}^M\alpha_{r,j}\kt{\bh_j},\quad r=(M+1),\ldots,\cbn_1.
\enddis
The aforesaid linearly independent vectors can be singled out by the {\em centred square}  
procedure reported in \S 5 of II because $\cQ_{\cS_1}$ is positive definite in the subspace 
spanned by $\kt{\bh_1},\ldots,\kt{\bh_M}$.  By so doing we shall find some KH zeros, but 
the KH matrix (with elements $\cI_{\bh_l,\bh_m}$) relevant to the set of vectors 
$\cB_M=\{\kt{\bh_1},\ldots,\kt{\bh_M}\}$ certainly is non-singular. On the contrary, 
the KH matrices relevant to the $(\cbn_1-M)$ 
sets of vectors $\cB_M\cup\kt{\bh_r}$, with $r=(M+1),\ldots,\cbn_1$, are singular and all the 
${\bh_r}$, with $r=(M+1),\ldots,\cbn_1$ are KH zeros. However, the locations of these zeros 
must be such that the configuration of $\cB_{M}$ is not that of a basic set, in the sense 
that the locations of the KH zeros is such that an "enlargement" of $\cB_M$ by the centred 
square procedure is possible whenever one could dispose of a set $\cS'_1\supset \cS_1$ (as 
it would happen with a larger limiting sphere). One concludes that the configuration of 
the KH zeros associated to $\cB_M$ 
is alike to that found in the previous case, \ie\ when $\cS_1$ contains $\cbn$ linearly 
independent vectors and a basic set is not found. 
Hence, only two cases are possible: either $\cS_1$ is large enough to contain a basic set or 
it is not. In the second case, the observed diffraction pattern does not allow to solve 
the phase problem. In the first case, it does. In fact, as already stressed, the singled-out 
basic set is also a basic set for the subtracted intensities $I_{\bh}$ and we can use all 
the results found in papers I and II for the case of X-ray scattering\footnote{
Strictly speaking, the determination of matrix $({\cal T})$ as reported in Appendix D of I is not 
possible in the case of neutron scattering. Hence, the matrix $({\cal R})$, instead 
of being evaluated by Eq. (I.2.41), must be evaluated by the first equality in Eq.~(I.2.45), 
\ie\ by inverting the KH matrix (with elements $I_{\bh_l,\bh_m}$) associated to the 
basic set of reflections.}. In particular, the 
knowledge of the basic set and of the further reflections where we have found a KH zero 
allow us to determine the associated {\em complete} set ($\cC$) and {\em complementary} 
set ($\cF$) of reflections. The first consists of the reflections differences of any two 
reflections of the basic set $\cB_{\cbn}$, and the second of the reflections (not contained 
in the complete set) that are differences of the reflections associated to the KH zeros 
with the reflections of the basic set. The recursive determination of 
the intensities relevant to the reflections not contained in $\cC\cup \cF$ proceeds 
along the lines described in I and one finds that also in the case of neutron 
scattering the full diffraction pattern is determined by the knowledge of the 
intensities associated to the finite set of reflections $\cC\cup \cF$. Moreover, 
each KH zero states that the added vector to $\cB_m$ is a linear combination of 
the vectors of $\cB_m$. The coefficients of the linear combination are obtained by 
solving the associated linear equations whose matrix of coefficients is the KH matrix 
associated to $\cB_m$. The scalar product with $\bra{\vd{\hj}}|$ yields a polynomial 
equations in the variables $e^{-i2\pi x_{\hj}}$ and $e^{-i2\pi y_{\hj}}$ (we recall that 
we are considering the case $D=2$). The system of these polynomial equations has 
$2\cbn$ common unimodular roots that are related to the positions of the $\cbn$ 
scattering centres as it is just specified. It is interesting to note that, by 
following the procedure reported in sect. 6 of II,  the aforesaid system of polynomial 
equations in two variables can be converted into a set of polynomial 
equations in a single variable, so as to make the solution of the problem simpler. 
\bigskip
\section{Conclusion}
The results reported in \S 3 and appendices A and B refer to the two-dimensional case. 
Their extension to the case $D=3$ is rather straightforward by following the lines 
illustrated in paper II. In particular, we write now $\delta_{\hj}$ as 
$(x_{\imath},y_{\imath,\jmath},z_{\imath,\jmath,\ell})$ where index 
$\imath$ labels the different projections of all the $\vd{\hj}$'s along ${\bf a}=
{\hat x}$, $\jmath$ the different projections of the $\vd{\hj}$'s that have the same 
${\hat x}$ projection along ${\bf b}={\hat y}$, and $\ell$ the different projections 
along ${\bf c}={\hat z}$ of the $\vd{\hj}$'s that have the same ${\hat x}$ and ${\hat y}$ 
projections. Then the set $\cI$ becomes 
\begin{equation}\label{6.1}
\cI\equiv\{(\imath,\jmath,\ell)\Bigl |\ 1\le \ell \le q_{\imath,\jmath},\ \ 1
\le \jmath \le p_{\imath}, \ \ 1\le \imath\le M\},
\end{equation}
with $\cbn=\sum_{\imath=1}^M\sum_{\jmath=1}^{p_{\imath}}q_{\imath,\jmath}$. Here 
label $\imath$ is assigned in such a way that $p_1\ge p_2\ge\cdots\ge p_M$. Then, for 
each $\imath$, label $\jmath$ is defined in such a way that $q_{\imath,1}\ge\cdots\ge 
q_{\imath,p_{\imath}}$, and similarly for $\ell$. After putting 
\begin{equation}\label{6.2}
\xi_{\imath}\equiv e^{-i2\pi  x_{\imath}},\quad \eta_{\imath,\jmath}\equiv e^{-i2\pi 
y_{\imath,\jmath}},\quad {\rm and }\quad \zeta_{\imath,\jmath,\ell}=e^{-i2\pi  
z_{\imath,\jmath,\ell}}
\end{equation}
a generic vector $\kt{\bh}$ with $\bh=(h,k,l)$ takes the form 
\begin{equation}\label{6.3}
\kt{h,k,l}=\sum_{\imath=1}^{M}\sum_{\jmath=1}^{p_{\imath}}\sum_{\ell=1}^{q_{\imath,\jmath}}
{\xi_{\imath}}^h {\eta_{\imath,\jmath}}^k {\zeta^l}_{\imath,\jmath,\ell}\kt{x_{\imath},
y_{\imath,\jmath},z_{\imath,\jmath,\ell} }.
\end{equation}
By the same analysis carried out in Appendix A, one finds that the basic set of 
reflections consists of the reflections $(h,k,l)$ determined by the conditions: 
$0\le l\le (q_{h+1,k+1}-1)$, $0\le k\le (p_{i+1}-1)$ and $0\le h\le (M-1)$, so that 
$\cB(\astar)$ coincides with $\cI$ shifted by (-1,-1,-1). The generic element 
$\cV_{\bh,\hj}$ of the associated $\cbn\times \cbn$ matrix $(\cV)$ now reads 
${\xi_{\imath}}^h {\eta_{\imath,\jmath}}^k {\zeta^l}_{\imath,\jmath,\ell}$ and, 
generalizing the procedure followed for obtaining Eq~(\ref{2.20}), one finds that 
its determinant is 
\begin{eqnarray}\label{6.4}
\det(\cV) &=&\Bigl [\prod_{\imath=1}^M \prod_{\jmath=1}^{p_{\imath}} 
\prod_{1\le s_1 < s_2\le q_{\imath,\jmath}}\Bigl (\zeta_{\imath,\jmath,s_2}-
\zeta_{\imath,\jmath,s_1}\bigr )\Bigr ]\cdot  \Bigl [\prod_{1\le \imath_1<\imath_2\le M}
\Bigl (\xi_{\imath_2}-\xi_{\imath_1}\bigr )^{\min(P_{\imath_2},P_{\imath_1})}\Bigr ]
\cdot{} \nonumber\\ 
& & {}\cdot     
\Bigl [\prod_{\imath=1}^M \prod_{1\le r_1<r_2\le p_{\imath}} 
\Bigl (\eta_{\imath,r_2}-\eta_{\imath,r_1}\bigr )^{\min (q_{\imath,r_2},
q_{\imath,r_2},)}\Bigr ],
\end{eqnarray}
where it has been put $P_{\imath}\equiv\sum_{\jmath=1}^{p_{\imath}}q_{\imath,\jmath}$ 
for $\imath=1,\ldots,M$. Since Eq.~(\ref{karlem}) holds also true in the 3D case, 
one concludes that the determinant of the KH matrix associated to the 3D principal 
basic set of reflections defined above has the following algebraic expression
\begin{eqnarray}\label{6.5}
\det\Bigl(\cD[\cB(\astar)]\Bigr) &=&\Bigl [\prod_{\imath=1}^M \prod_{\jmath=1}^{p_{\imath}} 
\prod_{1\le s_1 < s_2\le q_{\imath,\jmath}}\Bigl |\zeta_{\imath,\jmath,s_2}-\zeta_{\imath,
\jmath,s_1}\bigr |^2 \Bigr ]\cdot  \Bigl [\prod_{1\le \imath_1<\imath_2\le M}
\Bigl |\xi_{\imath_2}-\xi_{\imath_1}\bigr |^{2\min(P_{\imath_2},P_{\imath_1})}\Bigr ]
\cdot{} \nonumber\\
& & {}\cdot 
\Bigl [\prod_{\imath=1}^M \prod_{1\le r_1<r_2\le p_{\imath}} 
\Bigl |\eta_{\imath,r_2}-\eta_{\imath,r_1}\Bigr |^{2\min (q_{\imath,r_2},q_{\imath,r_2})}\Bigr ]
\cdot\Bigl[\prod_{\hj=1}^{\cbn}\nu_{\hj}\Bigr].
\end{eqnarray}

\noindent In conclusion, it has been shown that the algebraic approach applies both to X-ray and 
to neutron scattering. In this way it appears clear that the atomicity is the crucial 
assumption, while the positiveness of the scattering density is from a theoretical point 
of view not important. It only makes the search of the basic set faster, because the 
search can be carried through with the subtracted intensities $I_{\bh}$. On the contrary, 
in the case of neutrons, the procedure is slightly more involved, because one must first 
select the largest $\cS_1$ subset within the observed diffraction pattern and by these 
intensities  to evaluate the $\cJ_{\bh_r,\bh_s}$'s for all the $\bh_r$ and $\bh_s$ 
of $\cS_1$. Then, the search of a basic set is performed, as in the case of X-ray 
scattering, using the $\cJ_{\bh_r,\bh_s}$'s. Finally, after finding a basic set, 
one proceeds with the $I_{\bh}$'s and with the found basic set to reconstruct the 
full diffraction pattern and to determine the atomic positions as in the case of 
X-ray scattering. 

\vfill\eject
\appendix
\section {Principal basic sets for the 2-D case}
We show now that $\cZ^3_v$ contains different sets of $\cbn$ linearly independent 
vectors and we report the procedure for selecting one of these sets. Actually, 
this proof is immediately achieved by the procedures illustrated in I and II if 
we assume to know the quantities 
\begin{equation}\label{A1}
{\tilde {I}}_{\bh,\bk}\equiv\bra{\bh}\kt{\bk}=\sum_{\hj=1}^{\cbn}e^{i2\pi(\bh-\bk)\cdot\vd{j}}=
{\tilde {I}}_{\bh-\bk}
\end{equation}
[where the rhs follows from Eq.~(\ref{1.10c})], because the involved vectors belong 
to $\cZ^3_v$ and, therefore, belong to a Hilbert space so that the assumptions, made 
in I and II, are obeyed\footnote{The comparison of (\ref{A1}) with (\ref{1.5}) shows that 
the ${\tilde I}_{\bh-\bk}$'s can be considered as the scattering intensities relevant to 
a set of $\cbn$ scattering centres, located at $\vd{1},\ldots,\vd{\cbn}$, with positive 
charges all equal to one.}. We prefer however to proceed differently in order to make it 
evident that the geometry of the scattering centres determines the principal basic set 
of reflections. To this aim, we first observe that all the equations reported in \S 2 
hold also true if we restrict ourselves, for greater notational simplicity, to the 
case of a 2D space. We remark that, even though the locations of the $\cbn$ scattering 
centres are different from each other, it can happen that the distinct projections of 
the $\vd{\hj}$'s along one of the crystal axes are only $M$ with $M<\cbn$. Hence, we 
shall denote the distinct projections along ${\hat x}={\bf a}$ by $x_1,\dots,x_M$. 
Consider now those $\vd{\hj}$'s that have $x$-projections equal to, say, $x_1$. Since 
these $\vd{\hj}$'s are different, their projections along axis $y$ must differ among 
themselves and we shall denote their number by $m_1\ge 1$. In this way, the considered 
$\vd{\hj}$'s can be written as $[(x_1,y_1), (x_1,y_2),\ldots,(x_1,y_{m_1})]$ and, 
in general, we have 
\begin{equation}\label{2.1}
\Bigl\{\vd{1},\dots,\vd{\cbn}\Bigr\}=\Bigl\{(x_r,y_{r,s})\ |\ s=1,2,\ldots,m_r, \ 
r=1,\ldots,M,\Bigr\}
\end{equation}
Further, in labeling the different $x_r$, we choose the label in such a way that 
$m_1\ge m_2\ge\dots\ge m_M$. It is observed that points $(x_r,y_{rs})$ can be mapped 
into a subset $\cI$ of the $\cZ^2$ lattice defined either as 
\begin{equation}\label{latt}
 \cI\equiv\{(r,s)\Bigl |\  s-1,\ldots,m_r,\  r=1,\ldots,M\}
 \end{equation}
or as 
\begin{equation}\label{lattb}
 \cI\equiv\{(r,s)\Bigl|\ r=1,\ldots,\mu_s,\ \  s=1,\ldots,m_1\}
\end{equation} 
In the first case, $m_r$ is the number of points lying on the $r$th column, while in 
the second $\mu_s$ is the number of points lying on the $s$th row. In both cases, we 
have $\sum_{r=1}^M m_r=\sum_{s=1}^{m_1}\mu_s=\cbn$. Writing $\kt{\bh}$ as $\kt{h,k}$ 
with $h$ and $k$ integers, Eq.~(\ref{1.10b}) reads                    
\begdis
\kt{h,k}=\sum_{r=1}^{M}\sum_{s=1}^{m_r}e^{-i2\pi h x_r}e^{-i2\pi k y_{r,s}} \kt{x_r,y_{r,s}},
\enddis                                     
or, after putting
\begin{equation}\label{2.2}
\xi_r\equiv e^{-i2\pi  x_r}\quad {\rm and }\quad \eta_{r,s}\equiv e^{-i2\pi  y_{r,s}},
\end{equation}
as
\begin{equation}\label{2.3}
\kt{h,k}=\sum_{r=1}^{M}\sum_{s=1}^{m_r}{\xi_r}^h {\eta_{r,s}}^k \kt{x_r,y_{r,s}}.
\end{equation}
Consider now the polynomial 
\begin{equation}\label{2.4}
P_{_M}(z)=\prod_{r=1}^M(z-\xi_r)=\sum_{s=0}^M \alpha_{_M,s} z^s
\end{equation}
with
\begin{equation}\label{2.5}
\alpha_{_{M,s}}=(-1)^{M-s}\sum_{\scriptscriptstyle 1\le s_1< s_2\dots< s_{M-s}
\le M} \xi_{s_1}\xi_{s_2}\dots
\xi_{s_{M-s}},\quad s={0,\ldots,(M-1)},
\end{equation}
and $\alpha_{_{M,M}}=1.$ Eq.(\ref{2.4}) implies that $P_{_M}(\xi_r)=0$ for 
$r=1,\dots,M$. From Eq.~(\ref{2.4}) one gets 
\begin{equation}\label{2.6}
{\xi_r}^M=-\sum_{s=0}^{M-1} \alpha_{M,s} {\xi_r}^s\equiv \sum_{s=0}^{M-1} 
\beta_{M,M,s}{\xi_r}^s,\quad  r={1,\ldots,M}
\end{equation} 
and substituting these relations in $\kt{M,k}$ one obtains
\begin{equation}\label{2.7}
\kt{M,k}=-\sum_{p=0}^{M-1}\alpha_{M,p}\sum_{r=1}^{M}\sum_{s=1}^{m_r}{\xi_r}^p 
{\eta_{r,s}}^k \kt{x_r,y_{r,s}}=\sum_{s=0}^{M-1}\beta_{M,M,s}\kt{s,k},
\end{equation}
which shows that $\kt{M,k}$ is a linear combination of vectors $\kt{s,k}$ 
defined as 
\begin{equation}\label{2.7a}
\kt{s,k}\equiv \sum_{r=1}^{M}\sum_{p=1}^{m_r}{\xi_r}^s {\eta_{r,p}}^k\kt{x_r,y_{r,p}},
\quad s={0,\ldots,(M-1)}.
\end{equation}
Moreover, after multiplying both sides of Eq.(\ref{2.6}) by ${\xi_r}$ and using again 
Eq.(\ref{2.6}), one gets
\begin{equation}\label{2.7b}
{\xi_r}^{M+1}=-\sum_{s=0}^{M-2} \alpha_{_{M,s}} {\xi_r}^{s+1}-\alpha_{_{M,({M-1})}} 
{\xi_r}^M\equiv \sum_{s=0}^{M-1}\beta_{_{M,(M+1),s}} {\xi_r}^s,\quad r=1,\ldots,M,
\end{equation}
with 
\begin{eqnarray}\label{2.7c}
\beta_{_{M,(M+1),s}}  = \left\{ \begin{array}{ll}
\alpha_{_{M,(M-1)}}\alpha_{_{M,0}},&\textrm{if $s=0$,}\\
-\alpha_{_{M,s-1}}+\alpha_{_{M,M-1}}\alpha_{_{M,s}},&\textrm{if\   $s={1,\ldots,(M-1).}$}
\end{array}\right.
\end{eqnarray}
Eq.~(\ref{2.7b}) shows that ${\xi_r}^{M+1}$ also is a linear combination of ${\xi_r}^{0},
\dots,{\xi_r}^{M-1}$ with coefficients $\beta_{_{M,(M+1),s}}$ specified by 
Eq.~(\ref{2.7c}) and obtained by a recursive application of Eq.(\ref{2.6}). 
After dividing the equation  $P_{_M}(\xi_r)=0$ by ${\xi_r}^{-1}$, one obtains that 
\begdis
{\xi_r}^{-1}=-\alpha_{{M,0}}^{-1}\sum_{s=0}^{M-1} \alpha_{M,s+1} {\xi_r}^s\equiv 
\sum_{s=0}^{M-1} \beta_{_{M,-1,s}} {\xi_r}^s, \quad 
r={1,\ldots,M}. 
\enddis
Iterating the procedure one finds that 
\begin{equation}\label{2.8b}
{\xi_r}^{k}=\sum_{s=0}^{M-1} \beta_{M,k,s} {\xi_r}^s,\quad\quad r=1,\dots,M,
\quad\quad \forall k\ge M,\quad \forall k\le -1.
\end{equation}
>From these relations it follows that 
\begin{equation}\label{2.9}
\kt{h,k}=\sum_{s=0}^{M-1}\beta_{M,h,s}\kt{s,k}, \quad \forall h\ge M,\ 
\forall \ h\le -1.
\end{equation}
Consider now the polynomials
\begin{equation}\label{2.9a}
P_{m_r}(z)=\prod_{s=1}^{m_r}(z-\eta_{r,s}),\quad {\rm for}\quad r=1,\dots,M.
\end{equation}
By the same procedure, one proves that 
\begin{equation}\label{2.*}
{\eta_{r,s}}^{k}=\sum_{q=0}^{m_r-1} \beta_{m_r,k,q} {\eta_{r,s}}^q,\quad 
r=1,\dots,M,\ s=1,\dots,m_r, \ \forall  k\ge m_r,\ \forall \ k\le -1
\end{equation}
where coefficients $\beta_{m_r,k,q}$ are iteratively determined in terms of 
the coefficients that define the polynomials (\ref{2.9a}).  
We put now 
\begin{equation}\label{2.10}
\kt{x_r,k}\equiv \sum_{s=1}^{m_r}{\eta_{r,s}}^{k}\kt{x_r,y_{r,s}},\quad k=0,1,\ldots. 
\end{equation}
>From Eqs~(\ref{1.9a}) and (\ref{2.2}) it follows that 
\begin{equation}\label{2.10a}
\bra{x_s,k'}\kt{x_r,k}=\delta_{s,r}\sum_{s=1}^M {\eta_{r,s}}^{k-k'}
\end{equation}
and one concludes that vectors $\kt{x_r,k}$ and $\kt{x_s,k'}$ are linearly 
independent if $r\ne s$. The same happens for the vectors $\kt{x_r,0},\cdots$,$
\kt{x_r,(m_r-1)}$. This property immediately follows from Eq.~(\ref{2.10}), because 
the rhs of the equation involves $m_r$ linearly independent vectors at fixed $r$ and 
the matrix of coefficients ${\eta_{r,s}}^{k}$, with $k=0,\ldots,(m_r-1)$ and $s=1,
\ldots,m_r$, is a Vandermonde matrix with determinant equal to $\prod_{1\le p<q\le m_r}
(\eta_{r,q}-\eta_{r,p})$, which is certainly different from zero because 
the $\eta_{r,q}$'s are all different among themselves. Combining Eqs~(\ref{2.3}) 
and (\ref{2.10}) one finds that 
\begin{equation}\label{2.10b}
\kt{h,k}=\sum_{r=1}^{M}{\xi_r}^h \kt{x_r,k}.
\end{equation}
If $k=0$, all the vectors on the rhs are linearly independent as $r$ ranges from 1 to $M$. 
If we let $h$ range in $[0,\ldots,(M-1)]$, we have $M$ linear relations and the coefficient 
matrix is a Vandermonde matrix with determinant equal to $\Bigl[\prod_{1\le \ell < 
\jmath\le M}(\xi_{\jmath}-\xi_{\ell})\Bigr]\ne 0$. Thus, the linear independence of 
$\kt{x_1,0},\dots,\kt{x_M,0}$ ensures the linear independence of the vectors $\kt{0,0},
\ldots,\kt{(M-1),0}$. If $m_M>1$, by the same procedure and using the linear independence 
of the sets of vectors $\Bigl[\kt{x_1,k},\ldots,\kt{x_M,k}\Bigr]$ for $k=0,\ldots,(m_M-1)$, 
one finds that each of the following sets of vectors $\Bigl[\kt{0,\ell},\ldots,\kt{(M-1),
\ell}\Bigr]$, with $\ell=0,\ldots,(m_M-1)$, is a set of $M$ linearly independent vectors. 
Moreover, the vectors $\Bigl[\kt{0,\ell},\ldots,\kt{(M-1),\ell}\Bigr]$ are linearly 
independent from the vectors $[\kt{0,\jmath},\ldots,\kt{(M-1),\jmath}]$ if $0\le \ell 
\ne \jmath\le (m_M-1)$ due to Eq.~(\ref{2.10a}). Consider now the $m\times m$ Vandermonde 
matrix $\Bigl(\cV(m,\xi)\Bigr)$ with elements 
\begin{equation}\label{vander}
\cV_{j,k}(m,\xi)=\xi_j^{k-1}, \quad 1\le j,k\le m. 
\end{equation}
As far as $m\le M$, the aforesaid matrix is non-singular and endowed of an inverse 
denoted by $\Bigl(\cV^{-1}(m,\xi)\Bigr)$. Then, one can write
\begin{equation}\label{2.10c}
\kt{x_r,q}=\sum_{s=1}^M{\cV^{-1}}_{r,s}(M,\xi)\kt{(s-1),q},\ 1\le r\le M, \quad 0\le q 
\le (m_M-1).
\end{equation}
Assume now that $m_{M-1}>m_M$ and consider those values of $k$ such that $m_M\le k<
m_{M-1}$. Eq.~(\ref{2.3}) can be written by (\ref{2.*}) as 
\begin{eqnarray}
\lefteqn{\kt{h,k} = \sum_{r=1}^{M-1}\sum_{s=1}^{m_r}{\xi_r}^h {\eta_{r,s}}^k \kt{x_r,y_{r,s}}
+ \sum_{q=0}^{m_M-1}{\xi}^h \beta_{_{m_M},k,q} \kt{x_{_M},q} {} }\nonumber \\
   & = &{} \sum_{r=1}^{M-1}{\xi_r}^h\kt{x_r,k}+ \sum_{q=0}^{m_{_M}-1}{\xi_{_M}}^h 
\beta_{_{m_M},k,q} \kt{x_{_M},q} {}\nonumber 
\end{eqnarray}
that by Eq.~(\ref{2.10c}) becomes 
{\setlength\arraycolsep{2pt}
\begin{eqnarray}\label{2.16} 
\kt{h,k} & - & \sum_{q=0}^{m_{_M}-1}{\xi_{_M}}^h \beta_{{m_M},k,q}\sum_{s=1}^M{\cV^{-1}}_{r,s}
(M,\xi)\kt{(s-1),q} {}\nonumber \\
&=& {}  \sum_{r=1}^{M-1}{\cV}_{r,(h+1)}(M-1,\xi)\kt{x_r,k}, \quad m_M\le k\le(m_{M-1}-1). 
\end{eqnarray}}
\noindent The vectors $\kt{x_r,k}$ present on the rhs of (\ref{2.16}) are linearly independent 
for $k=0,\dots,(M-1)$ and the matrix $\Bigl(\cV(M-1,\xi)\Bigr)$ is non singular. Thus, the 
vectors on the lhs are also linearly independent for $0\le h\le (M-1)$ and $m_M\le k\le 
(m_{M-1}-1)$. The vectors within the sum present on the lhs of (\ref{2.16}) were already 
shown to be linearly independent because they are characterized by $q$ values ranging 
in $[0, (m_M-1)]$. One concludes that the vectors $\kt{h,k}$ with $0\le h\le (M-1)$ and 
$m_M\le k\le (m_{M-1}-1)$ are linearly independent. In this way, we have proven that the 
vectors $\kt{h,k}$ with $0\le k \le(m_{M-1}-1)$ are linearly independent if $0\le h\le 
(\mu_{k+1}-1)$, the integers $\mu_k$ being defined by Eq.~(\ref{lattb}). If $m_{M-1}=m_{M}$, 
the inequality $m_{M}\le k < m_{M-1}$ is never verified and it must be substituted with 
$m_{M-1}\le k< m_{M-2}$ provided $m_{M-2}>m_{M-1}(=m_M)$. In this way, step by step, 
by the procedure just reported one shows that the vectors linearly independent are 
{\setlength\arraycolsep{2pt}
\begin{eqnarray}\label{baset}
\cB(\astar) & \equiv & \Bigl\{\,\kt{h,k}\,\Bigl|\quad h=0,\ldots,(\mu_{k}-1), \quad 
k=0,\ldots,(m_1-1)\Bigr\}{}\nonumber \\
&=& {} \Bigl\{\,\kt{h,k}\,\Bigl| \quad k=0,\ldots,(m_h-1), \quad h=0,\ldots,(M-1)\Bigr\}.
\end{eqnarray}}
The corresponding set of $(h,k)$ is nothing else that set $\cI$, specified by 
Eq.~(\ref{lattb}) and shifted by (-1,-1). Hence it is fully specified by the geometry 
of the $\vd{\hj}$ values, once these values are mapped into $\cI$. The set $\cB(\astar)$ 
consists of $\cbn$ points. It will be referred to as the {\it principal basic} set of 
vectors along reciprocal crystallographic axis $\astar$: {\it basic} because $\cB(\astar)$ 
determines a complete basis of $\cHr$, linearly related to that formed by 
$\kt{\vd{1}}$,..,$\kt{\vd{\cbn}}$ and used to define $\cHr$, and {\it principal} (along 
$\astar$) because $\cB(\astar)$ has the largest extension along $\astar$ since, at each 
step of the procedure, we tried to include the largest number of reflections $(h,k)$ 
lying on the rows parallel to $\astar$. Any possible confusion being avoided by the context, 
the set of reflections $(h,q)$ with $h$ an $k$ obeying to the constraints 
specified in Eq.(\ref{baset}) will also be denoted by $\cB(\astar)$ and named 
principal basic set of reflections. Papers I and II showed the existence of less 
elongated basic sets as well as the procedures for singling them out. These 
procedures are based on an "enlargement" method dictated by the Karle-Hauptman 
zeros found during the basic set search. On the contrary, the procedure illustrated 
above only bases on the geometry of the locations of the scattering centres. 

\section {Generalized Vandermonde determinant}
In Appendix A we showed that the vectors $\kt{h,k}$, with $h$ and $k$ obeying 
Eq.~(\ref{baset}), are linearly independent because they form the principal 
basic set of vectors along $\astar$. Then, the associated matrix $(\cV)$ with its 
elements defined by Eq. (\ref{2.3}) must be non-singular. The analytical 
expression of the determinant of this matrix $(\cV)$ is remarkably simple. 
To get this expression, we note that the full expression of $(\cV)$ is 
\begin{equation}\label{vandermatr}{\setlength\arraycolsep{1pt}
(\cV)={\tiny { \left( 
\begin{array}{ccccccccccc} 
\tom{1}{1}{0}{0}& \tom{1}{1}{0}{1}&\cdot&\tom{1}{1}{0}{m_1-1}
&\tom{1}{1}{1}{0}&\cdot &\tom{1}{1}{1}{m_2-1} &\cdot 
&\tom{1}{1}{^{M-1}}{0}& \cdot & \tom{1}{1}{^{M-1}}{m_{_M}-1}\\
\tom{1}{2}{0}{0}& \tom{1}{2}{0}{1}&\cdot&\tom{1}{2}{0}{m_1-1}
&\tom{1}{2}{1}{0}&\cdot &\tom{1}{2}{1}{m_2-1} &\cdot 
&\tom{1}{2}{^{M-1}}{0}& \cdot &\tom{1}{2}{^{M-1}}{m_{_M}-1}\\
\vdots&\vdots&\vdots&\vdots&\vdots&\vdots&\vdots&\vdots&\vdots&\vdots&
\vdots\\
\tom{1}{{m_1}}{0}{0}& \tom{1}{m_1}{0}{1}&\cdot&\tom{1}{m_1}{0}{m_1-1}
&\tom{1}{m_1}{1}{0}&\cdot &\tom{1}{m_1}{1}{m_2-1} &\cdot 
&\tom{1}{m_1}{^{M-1}}{0}& \cdot &\tom{1}{m_1}{^{M-1}}{m_{_M}-1}\\
\tom{2}{1}{0}{0}& \tom{2}{1}{0}{1}&\cdot&\tom{2}{1}{0}{m_1-1}&\tom{2}{1}{1}{0} 
&\cdot &\tom{2}{1}{1}{m_2-1} &\cdot &\tom{2}{1}{^{M-1}}{0}& \cdot &
\tom{2}{1}{^{M-1}}{m_{_M}-1}\\
\tom{2}{2}{0}{0}& \tom{2}{2}{0}{1}&\cdot&\tom{2}{2}{0}{m_1-1}&\tom{2}{2}{1}{0} 
&\cdot &\tom{2}{2}{1}{m_2-1} &\cdot &\tom{2}{2}{^{M-1}}{0}& \cdot &
\tom{2}{2}{^{M-1}}{m_{_M}-1}\\
\vdots&\vdots&\vdots&\vdots&\vdots&\vdots&\vdots&\vdots&\vdots&\vdots&
\vdots\\
\tom{2}{m_2}{0}{0}& \tom{2}{m_2}{0}{1}&\cdot&\tom{2}{m_2}{0}{m_1-1}
&\tom{2}{m_2}{1}{0} 
&\cdot &\tom{2}{m_2}{1}{m_2-1} &\cdot &\tom{2}{m_2}{^{M-1}}{0}& \cdot &
\tom{2}{m_2}{^{M-1}}{m_{_M}-1}\\
\vdots&\vdots&\vdots&\vdots&\vdots&\vdots&\vdots&\vdots&\vdots&\vdots&
\vdots\\
\vdots&\vdots&\vdots&\vdots&\vdots&\vdots&\vdots&\vdots&\vdots&\vdots&
\vdots\\
\tom{{_{M}}}{{_{1}}}{0}{0}& \tom{_M}{{_1}}{0}{1}&\cdot&\tom{_{M}}{{_{1}}}{0}{{m_{1}-1}}
&\tom{{_{M}}}{{_{1}}}{1}{0} &\cdot &\tom{{_M}}{{_1}}{1}{{m_2-1}} &\cdot 
&\tom{{_{M}}}{{_{1}}}{{^{M-1}}}{0}& \cdot &\tom{{_M}}{{_1}}{{^{M-1}}}{{m_{_{M}}-1}}\\
\tom{{_{M}}}{{_{2}}}{0}{0}& \tom{{_M}}{{_2}}{0}{1}&\cdot&\tom{{_M}}{{_{2}}}{0}{{m_{1}-1}}
&\tom{{_{M}}}{{_{2}}}{1}{0} &\cdot &\tom{{_{M}}}{{_2}}{1}{{m_2-1}} &\cdot 
&\tom{{_{M}}}{{_{2}}}{{^{M-1}}}{0}& \cdot &\tom{{_M}}{{_2}}{{^{M-1}}}{{m_{_{M}}-1}}\\
\vdots&\vdots&\vdots&\vdots&\vdots&\vdots&\vdots&\vdots&\vdots&\vdots&
\vdots\\
\tom{{_{M}}}{{m_{_{M}}}}{0}{0}& \tom{{_{M}}}{{_{m_{_{M}}}}}{0}{1}&\cdot&\tom{_{M}}{{_{m_{_{M}}}}}{0}{m_1-1}
&\tom{{_{M}}}{{_{m_{_{M}}}}}{1}{0} &\cdot &\tom{{_{M}}}{{_{m_{_{M}}}}}{1}{m_2-1} &\cdot 
&\tom{{_{M}}}{{_{m_{_{M}}}}}{^{M-1}}{0}& \cdot &\tom{{_{M}}}{{_{m_{_{M}}}}}{^{M-1}}{m_{_{M}}-1}
\end{array}\right ) {}} }}   
\end{equation}
The rows of $(\cV)$ correspond to $(r,s)$ with $s=1,\dots,m_r$ and $r=1,\dots,M$
and the columns to $(p,q)$ with $q=0,\dots,(m_{p+1}-1)$ and $p=0,\dots,(M-1)$. 
The determinant of $(\cV)$ is a homogeneous polynomial in variables $\{\xi\}$ and 
$\{\eta\}$, because each term of $\det(\cV)$ has degree $Q$ with respect to variables 
$\{\eta\}$ and degree $P$  in the $\{\xi\}$'s. $Q$ and $P$ respectively are 
{\setlength\arraycolsep{2pt}
\begin{eqnarray}\label{2.18a}
Q &  = & [0+1+\cdots+(m_1-1)]+[0+1+\cdots+(m_2-1)]+\dots+{} \nonumber  \\
&  & {}+[0+1+\cdots+(m_{_{M}}-1)]=\sum_{k=1}^M {m_k}^2/2-\cbn/2, {}
\end{eqnarray}}
and 
\begin{equation}\label{2.18b}
P\equiv 0\cdot m_1+1\cdot m_2+\dots+(M-1)\cdot m_{_{M}}
\end{equation}
If, whatever $i$,  $\eta_{i,j}=\eta_{i,l}$ with $j\ne l$, two rows of $(\cV)$ are 
equal and the determinant will be equal to zero.  Thus, one can write 
\begin{equation}\label{2.19a}
\det(\cV)=\biggl [\prod_{i=1}^M\prod_{1\le j_1<j_2\le m_i} 
\bigl (\eta_{i,j_2}-\eta_{i,j_1}\bigr )\biggr ]\cdot \cR(\{\xi\},\{\eta\}).
\end{equation}
Since the total degree of the expression inside the square brackets in (\ref{2.19a}) 
is $Q$, one concludes that $\cR$ does not depend on variables  $\{\eta\}$ so 
that $\cR(\{\xi\},\{\eta\})=\cR(\{\xi\})$, and $\cR(\{\xi\})$ must be a 
polynomial of degree $P$. 
Evaluate now $\det(\cV)$ by considering the $(m_1\times m_1)$ 
minors contained in the first $m_1$ rows of $(\cV)$. From each column of the 
considered minor one first factorizes $\xi_1$ to an appropriate power. 
In this way we are left with an  $(m_1\times m_1)$  Vandermonde-like matrix 
with elements equal to $\eta_{1,j}^p$. It is remarked that, whenever $m_1>m_2$, 
a minor is non singular only if it contains the $(m_2+1), (m_2+2),\ldots,m_1$th 
columns of $(\cV)$ and, therefore, each complementary minor will not 
present terms $\eta_{r,j}^p$ with $2\le r \le M$ and $m_2\le j\le (m_1-1)$. 
Assume now that $\xi_1=\xi_j$, $j$ being a particular value such that 
$1<j\le M$. We have $m_1\ge m_j$. We can evaluate $\det(\cV)$ 
by considering all the minors containing the rows $(1,r)$ with 
$1\le r\le m_1$ and the rows $(j,s)$ with $1\le s\le m_j$. Having assumed 
that $\xi_1=\xi_j$, each column of one of these minors factorizes ${\xi_1}^p$ 
with $p$ in $[0,\dots,(M-1)]$ depending on the considered column. Thus, we are 
left with an $(m_1+m_j)\times (m_1+m_j)$ Vandermonde matrix whose elements 
are ${\eta_{1,p}}^r$ or ${\eta_{j,q}}^s$ with $1\le p\le m_1$,  $1,\le q \le m_j$, 
while $r$ and $s$ belong to $[0,\dots,(m_1-1)]$. The rank of this matrix at most 
is equal to $m_1$ and, therefore, its determinant will have a zero 
of order $m_j=\min(m_1,m_j)$. Hence, $\det(\cV)$ evaluated by this procedure 
will have a zero, related to the fact that $\xi_1=\xi_j$, of order $m_j$. 
Assume now that $\xi_2=\xi_j$ for a particular $j$ such that 
$2<j\le M$. Before repeating the reasoning made in the case 
$\xi_1=\xi_j$, we imagine of having developed $\det(\cV)$ with 
respect to the minors contained in the first $m_1$ rows. 
As noted above, each complementary minor will no longer contain 
the columns with exponents $m_2,\dots,(m_1-1)$. Each of these 
complementary minor can be developed by considering its 
$(m_2+m_j)\times (m_2+m_j)$ minors contained in the rows presenting 
the factors $\xi_2$ and $\xi_j$. By the same reasoning made above 
for the case $\xi_1=\xi_j$, one concludes that each $(m_2+m_j)\times 
(m_2+m_j)$ minor has a rank at most equal to $m_2$ and, therefore 
$\det(\cV)$ must have a zero of order at least equal to $m_j=\min(m_2,m_j)$ 
when $\xi_2=\xi_j$. One concludes that the zero of $\det(\cV)$ 
is at least of the order $\min(m_,m_j)$ when 
$\xi_i=\xi_j$ and one can write that 
\begin{equation}\label{2.19b}
\det(\cV) = \biggl[\prod_{1\le l_1<l_2\le M}\Bigl(\xi_{l_2}-\xi_{l_2}\Bigr)
^{\min(m_{l_2},m_{l_1})}\biggr]\cdot \cR_1(\{\xi\},\{\eta\}).
\end{equation}
The degree of the $\xi$-polynomial inside square brackets is equal to $P$, 
so that $\cR_1$ is a polynomial of the only variables $\{\eta\}$. Thus, combining 
Eq.(\ref{2.19a}) with (\ref{2.19b}), one finds that 
\begin{equation}\label{64a}
\det(\cV) =\cR_0 \biggl[\prod_{1\le l_1<l_2\le M}\Bigl(\xi_{l_2}-\xi_{l_1}\Bigr)
^{\min(m_{l_2},m_{l_1})}\biggr] \biggl [\prod_{i=1}^M\prod_{1\le j_1<j_2\le m_i} 
\bigl (\eta_{i,j_2}-\eta_{i,j_1}\bigr )\biggr ],
\end{equation}
where $\cR_0$ is a simple constant, eventually dependent on the dimensionality 
of $(\cV)$. Comparing the "diagonal" term $\prod_{l=1}^{M} 
({\xi_l}^{l-1})^{m_l}\prod_{i=1}^M\prod_{j=1}^{m_i} {\eta_{ij}}^{j-1}$ 
resulting from the calculation of $\det(\cV)$, starting from the explicit expression 
of $(\cV)$, with the corresponding term obtained developing the products present in 
(\ref{64a}), one finds that $\cR_0=1$. Thus, the 
determinant of matrix $(\cV)$ is 
\begin{equation}\label{2.20}
\det(\cV) =\biggl[\prod_{1\le l_1<l_2\le M}\Bigl(\xi_{l_2}-\xi_{l_1}\Bigr)
^{\min(m_{l_2},m_{l_1})}\biggr] \biggl [\prod_{i=1}^M\prod_{1\le j_1<j_2\le m_i} 
\bigl (\eta_{i,j_2}-\eta_{i,j_1}\bigr )\biggr ].
\end{equation}
By this result it is possible to get the algebraic expression of the determinant 
of the Karle-Hauptman matrix $\Bigl(\cD[\cB(\astar)]\Bigr)$ associated to the 
principal basic set of vectors $\cB(\astar)$. To this aim, we denote the vectors
$\kt{\bh}\in \cB(\astar)$ as $\kt{\bh_{\ell}}$ with $\ell=1,\ldots,\cbn$. The $(\imath,\ell)$ 
element of $\Bigl(\cD[\cB(\astar)]\Bigr)$ is defined as $\cD_{\imath,\ell}[\cB(\astar)]=
I_{\bh_{\imath}-\bh_{\ell}}$. Then, by Eq.~(\ref{1.11}), one finds that
\begin{equation}\label{karle}
\cD_{\imath,\ell}[\cB(\astar)]=I_{\bh_{\imath}-\bh_{\ell}}=\bra{\bh_{\imath}}|\cQ
\kt{\bh_{\ell}}=\sum_{\hj=1}^{\cbn}e^{i2\pi \vd{\hj}\cdot\bh_{\ell}} \nu_{\hj}
e^{-i2\pi \vd{\hj}\cdot\bh_{\imath}},\ \ \imath,\ell=1,\ldots,\cbn.
\end{equation}
These can be written, using Eq.s~(\ref{5.1}) and (\ref{5.2}), as 
\begin{equation}\label{karlem}
\Bigl(\cD[\cB(\astar)]\Bigr) = \Bigl(\cV^{\dag}\Bigr)\cdot\Bigl(v\Bigr)\cdot\Bigl(\cV\Bigr)
\end{equation}
It follows that the determinant of the matrix on the lhs is the product of the determinants 
of the matrices present in the rhs. Hence, by Eq.~(\ref{2.20}), 
the determinant of the KH matrix associated to $\cB(\astar)$ is 
\begin{equation}\label{detkh}
\det\Bigl(\cD[\cB(\astar)]\Bigr) = \biggl[\prod_{\hj=1}^{\cbn}\nu_{\hj}\biggr]
\biggl[\prod_{1\le \imath<\jmath\le M}\Bigl|\xi_{\jmath}-\xi_{\imath}
\Bigr|^{2\min(m_{\jmath},m_{\imath})}\biggr] 
\biggl [\prod_{i=1}^M\prod_{1\le \jmath<\ell\le m_i} 
\bigl |\eta_{i,\ell}-\eta_{i,\jmath}\bigr |^2\biggr ].
\end{equation}
This expression applies both to X-ray and to neutrons. Its value is certainly 
different from zero. It is striclty positive in the first case while,  
in the second case, its sign depends on the sign of the first factor related 
to the product of the charges of the $\cbn$ scattering centres. 

\bigskip

\section*{References} 
\reference{Avrami, M.  {\em  Phys. Rev.} {\bf 50}, {300}, (1938).}
\reference{Buerger, M.J. {\em  Crystal-Structure analysis}, New York: Wiley, (1960).}
\reference{Cervellino, A. and Ciccariello, S. {\em  J. Phys. A: Math. Gen. {\bf 34}, 731, (2001).}}
\reference{Cervellino, A. and Ciccariello, S. {\em  Z. Kristall. {\bf 214}, 739, (1999).}}
\reference{Cervellino, A. and Ciccariello, S. {\em  Riv. Nuovo Cimento {\bf 19}/8, 1, (1996).}}
\reference{Fischer, K.F. and Pilz, K. {\em  Acta Cryst. {\rm A} {\bf 53}, 475, (1997).}}
\reference{Gantmacher, F.R. {\em  Th\'eorie des matrices}, Paris: Dunod. Vol. I, (1966).} 
\reference{Goedkoop, J.B. {\em  Acta Cryst. {\bf 3}, {374}, (1950).}}
\reference{Hauptman, H.A. {\em  Rep. Prog. Phys. {\bf 54},  {1427}, ({1991}).}}
\reference{Hauptman, H.A. {\em  Acta Cryst. A{\bf 32},  {877}, ({1976}).}}
\reference{Hauptman, H.A. and Langs, D.A. {\em  Acta Cryst. A{\bf 59},  {250}, ({2003}).}}
\reference{Karle, J. and Hauptman, H.A. {\em  Acta Cryst. {\bf 3}, 181, (1950).}}
\reference{Messiah, A. {\em  M\'ecanique quantique, } Paris: Dunod. Vol. I, (1959).}
\reference{Navaza, A. and Navaza, J. {\em  Acta Cryst. A {\bf 48}, 695, (1992).}}
\reference{Navaza, J. and Silva, A.M. {\em Acta Cryst. A {\bf 35}, 266, (1979).}}
\reference{Ott, H. {\em  Z. Kristall. {\bf 66},  {136}, (1927).}}
\reference{Patterson, A.L. {\em  Phys. Rev. {\bf 55}, 682, (1939).}}
\reference{Pilz, K. and Fischer, K.F. {\em  Z. Kristall. {\bf 215}, 640, (2000).}}
\reference{Rothbauer, R. {\em  Z. Kristall. {\bf 209}, 578, (1994).}}
\reference{Silva, A.M. and Navaza, J. {\em  Acta Cryst. {\rm A} {\bf 37}, 658, (1981).}}

\end{document}